\documentclass[%
 aip,
 amsmath,amssymb,
 reprint,%
]{revtex4-1}

\usepackage{graphicx}
\usepackage{dcolumn}
\usepackage{bm}

\usepackage[utf8]{inputenc}
\usepackage[T1]{fontenc}
\usepackage{mathptmx}
\usepackage{etoolbox}

\usepackage{xcolor}
\usepackage{subfigure}
\usepackage{tikz}
\usetikzlibrary{datavisualization}
\usetikzlibrary{plotmarks}
\usetikzlibrary{graphs}
\usetikzlibrary{shapes.geometric}
\usetikzlibrary{arrows,decorations.pathmorphing,backgrounds,positioning,fit,petri}
\usepackage{pgfplots}

\usepackage{soul}
\pgfplotsset{compat=newest}
\pgfplotsset{plot coordinates/math parser=false}

\pgfplotsset{compat=1.3,
	every axis legend/.style={
	y tick label style={/pgf/number format/1000 sep=},					
	x tick label style={/pgf/number format/1000 sep=},
	}}
\usetikzlibrary{arrows}

\makeatletter
\def\@email#1#2{%
 \endgroup
 \patchcmd{\titleblock@produce}
  {\frontmatter@RRAPformat}
  {\frontmatter@RRAPformat{\produce@RRAP{*#1\href{mailto:#2}{#2}}}\frontmatter@RRAPformat}
  {}{}
}%
\makeatother
\begin{document}

\preprint{AIP/123-QED}

\title[]{Chiral and directional optical emission from a dipole source coupled to a helical plasmonic antenna}

\author{Lilli Kuen}
\affiliation{Zuse Institute Berlin, Takustraße 7, 14195 Berlin, Germany}
\affiliation{JCMwave GmbH, Bolivarallee 22, 14050 Berlin, Germany}

\author{Lorenz Löffler}
\affiliation{Experimental Physics 5, Institute of Physics, University of Würzburg, Germany}

\author{Aleksei Tsarapkin}
\affiliation{Ferdinand-Braun-Institut, Leibniz-Institut für Höchstfrequenztechnik,  12489 Berlin, Germany}

\author{Lin Zschiedrich}
\affiliation{JCMwave GmbH, Bolivarallee 22, 14050 Berlin, Germany}

\author{Thorsten Feichtner}
\affiliation{Experimental Physics 5, Institute of Physics, University of Würzburg, Germany}

\author{Sven Burger}
\affiliation{Zuse Institute Berlin, Takustraße 7, 14195 Berlin, Germany}
\affiliation{JCMwave GmbH, Bolivarallee 22, 14050 Berlin, Germany}

\author{Katja Höflich}
\affiliation{Ferdinand-Braun-Institut, Leibniz-Institut für Höchstfrequenztechnik,  12489 Berlin, Germany}

\begin{abstract}
Plasmonic antennas with helical geometry are capable transducers between linearly polarized dipole emission and purely circular polarized far-fields.
Besides large Purcell enhancements they possess a wide tunability due to the geometry dependence of their resonant modes.
Here, the coupling of a dipole emitter embedded in a thin film to plasmonic single and double helices is numerically studied.
Using a higher-order finite element method (FEM) the wavelength dependent Purcell enhancement of a dipole with different positions and orientations is calculated and the far-fields with respect to their chirality and radiation patterns are analyzed. 
Both single and double helices demonstrate highly directional and circularly polarized far-fields for resonant excitation but with significantly improved directional radiation for the case of double helices.
\end{abstract}

\maketitle

Polarization is an important basic property of light related to the spin degree of freedom of its fundamental quanta, the photons. 
Right and left circularly polarized light corresponds to photons with spin $+1$ and $-1$, respectively. 
Accordingly, controlling the degree of (circular) polarization allows the realization of different photonic quantum information protocols~\cite{Flamini2019}. 
Such protocols rely on the interference of single indistinguishable photons emitted by one or several non-classical light sources, so-called quantum emitters.
While numerous advanced but macroscopic devices exist for photon polarization control, efficient implementation in future devices requires a nanoscale solution that combines strong light-matter interaction with a handle for setting or determining the polarization state~\cite{Moody2022}.

In this regard, metallic nanostructures are advantageous since their free-electron gas can be excited into collective movement by incident light. 
The associated quasi-particle describing light coupled to electron density oscillations is named plasmon-polariton and features electromagnetic near-fields at the metal surface.
Accordingly, the resonant excitation of a well-designed plasmonic antenna can exhibit confinement of optical fields several orders below the diffraction limit leading to extreme near-field concentration~\cite{Maier2007}.  
Therefore, plasmonic antennas provide an increased number of radiative decay channels for any dipolar emitter located in these fields leading to a correspondingly increased rate of spontaneous emission.
This is captured in Purcell's formula~\cite{purcell_spontaneous_1946}: $F= 3/(4\pi^2)(\lambda/n_{\text{m}})^3Q/V$, by the small mode volume.
Here, $F$ denotes the fluorophore's emission enhancement factor at a vacuum wavelength $\lambda$ when placed in a resonators field maximum with quality factor $Q$ and mode volume $V/(\lambda/n_{\text{m}})^3$, where the emitter resides in a medium of refractive index $n_{\text{m}}$.
This strongly enhanced light-matter interaction has already been used for sensing down to the single-molecule level e.g.~in Raman scattering~\cite{Kneipp1997} or molecular fluorescence~\cite{Kuhn2006, Kinkhabwala2009}.

Since plasmonic resonances are comparably broad and the internal quantum efficiency of the coupled emitter can approach unity, plasmonic antennas provide an exciting playground for all types of quantum emitters and their application as super-bright single photon sources~\cite{Novotny2011, Biagioni2012, Koenderink2017}.
Plasmonic antennas on the other hand can be also designed to couple to specific radiating modes~\cite{Feichtner2017}, and, therefore, e.g.~achieve directional emission of quantum emitters of different types~\cite{Taminiau2008, Curto2010} which is one of the main requirements for possible on-chip integration in a photonic architecture~\cite{Moody2022}.

Helical plasmonic geometries enable strong chiroptical interaction.
Arrays of helices were demonstrated to be transparent to one handedness of the incident light but reflective for the other~\cite{Gansel2009}. 
Similarly strong responses were shown for single helices made from silver \cite{hoflich2019resonant} approaching the theoretical maximum of chiroptical interaction~\cite{Fernandez-Corbaton2016, Garcia-Santiago2022}.
Hence, combining a helical plasmonic antenna with a (dipolar) quantum emitter would be an exciting opportunity to control the state of polarization at the nanoscale, i.e.~control the spin degree of freedom for single photons on a chip. 
First encouraging experimental results were achieved by creating a small slit at the bottom of a helical antenna mimicking a dipolar excitation source~\cite{Wang2019}.

Here, we numerically study the coupling of a linearly polarized dipolar emitter to two types of chiral plasmonic antennas: a single helix and a double helix.
The emitter resides in an hexagonal Boron nitride (hBN) layer, with hBN being one of the prime emerging platforms for bright and stable quantum emitters~\cite{Aharonovich2022}.
For the single helix, a semi-analytical model is used to assess resonance positions and far-field patterns.
The FEM modeling then treats the full scenarios of a linearly polarized dipole coupled to the helices to compute Purcell factors and detailed far-field emission characteristics.
Efficient coupling of excited dipoles can be achieved for both helix types leading to circularly polarized directional emission along the helix axis.

\begin{figure}
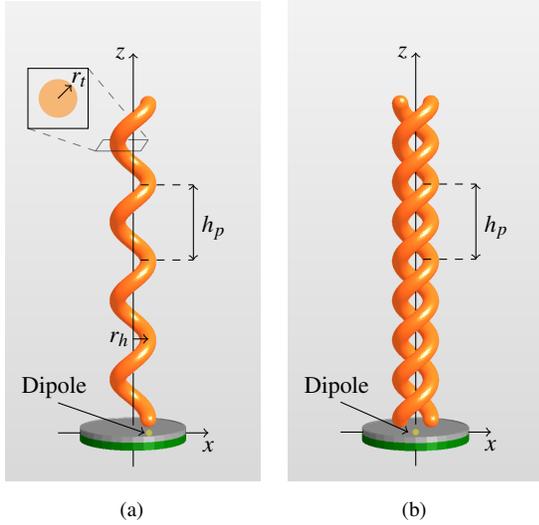

\centering
    \subfigure[]{\input{Bilder/FIG_1/Anordnung_Single_Helix}\label{fig:HelixAufbaua}}
    \subfigure[]{\input{Bilder/FIG_1/Anordnung_Double_Helix}\label{fig:HelixAufbaub}}
    \caption{Sketch of a single (a) and double (b) plasmonic helices with dipole emitter, pitch height $h_p$, helix radius $r_h$ and tube radius $r_t$.}
    \label{fig:HelixAufbau}
\end{figure}

{\it Theoretical Background and Implementation} -- 
We investigate chiral plasmonic resonators made from silver~\cite{johnson1972optical} similarily to~\cite{hoflich2019resonant} in the near-infrared regime arranged as single or double helices on a substrate (see Fig.~\ref{fig:HelixAufbau}).
The single helix is defined by its radius $r_h=60\,\text{nm}$, pitch height $h_p = 310\,\text{nm}$, amount of pitches $p$ and wire radius $r_t = 32\, \text{nm}$, with right-handed orientation as depicted in Fig.~\ref{fig:HelixAufbaua}.
The double helix with four pitches combines two of these single helices, one rotated by 180 degrees along the helix axis, see Fig.~\ref{fig:HelixAufbaub}.
The substrate consists of $20\,\text{nm}$ hBN~\cite{lee2019refractive} on $20\,\text{nm}$ Glass (BK7)~\cite{SchottBK7}. 
The helix ends dip $2\,\text{nm}$ into the hBN in order to realize a finite contact area between metal and substrate, and to obtain smooth meshing.
The entire setup is analyzed using a higher-order finite element method (FEM), implemented in the solver JCMsuite~\cite{pomplun2007adaptive}.  
The excitation is realized by a point source, described by a current density $\mathbf{J}(\mathbf{r},t)$.
For $\mathbf{r} \to \mathbf{r_0}$ with the position vector $\mathbf{r}$ and the position of the point emitter $\mathbf{r_0}$, the electric field diverges, which may affect the convergence of numerical results towards the exact solution.  
This problem is solved by a subtraction approach \cite{zschiedrich2013numerical}.
The Purcell enhancement is computed as the ratio between the dipole emission in the full system compared to emission in bulk hBN. 
This provides information about how much dipole power is coupled to the adjacent structures.
Due to the strong near-fields of resonant plasmonic structures, the radiative decay rate enhancement of dipolar emitters is extremely position-sensitive.
In addition, if the emitter comes too close, quenching will occur~\cite{Anger2006, Marquier2017}. 
Hence, careful modelling of dipole position, geometries and materials is required. 

With the vectorial three-dimensional field solution of the scattering simulation, the far-field can be calculated via Fourier transformation (cf. Supporting Information for more details). 
The degree of circular polarization in the far-field is retrieved to identify helix geometries able to transform linearly polarized dipole emission into circularly polarized light. 
Let $w_\textrm{RCP}$ and $w_\textrm{LCP}$ represent the amplitudes of the right-circularly polarized (RCP) and left-circularly polarized (LCP) components in the far-field in a given direction (for details see Supporting Information).
The degree of circular polarization $G$ is then defined as: 
\begin{equation}
    G = \frac{w_\textrm{LCP} - w_\textrm{RCP}}{w_\textrm{LCP} + w_\textrm{RCP}},
\end{equation}
which indicates in the range $[-1, 1]$ the degree of circular polarization from fully RCP to fully LCP. 
Technologically relevant is a directive large intensity of the far-field together with a high degree of polarization.
This is represented by the electric chirality density $ \chi_e$, given as \cite{gutsche2016time}
\begin{equation}
    \chi_e = \frac{1}{8} \left( \mathbf{D}^* \cdot \left( \nabla \times \mathbf{E} \right) + \mathbf{E} \cdot \left( \nabla \times \mathbf{D}^* \right) \right),
\end{equation}
where negative values correspond to RCP and positive  values to LCP light. 

{\it Numerical Investigation of a Single Helix} --
\begin{figure*}
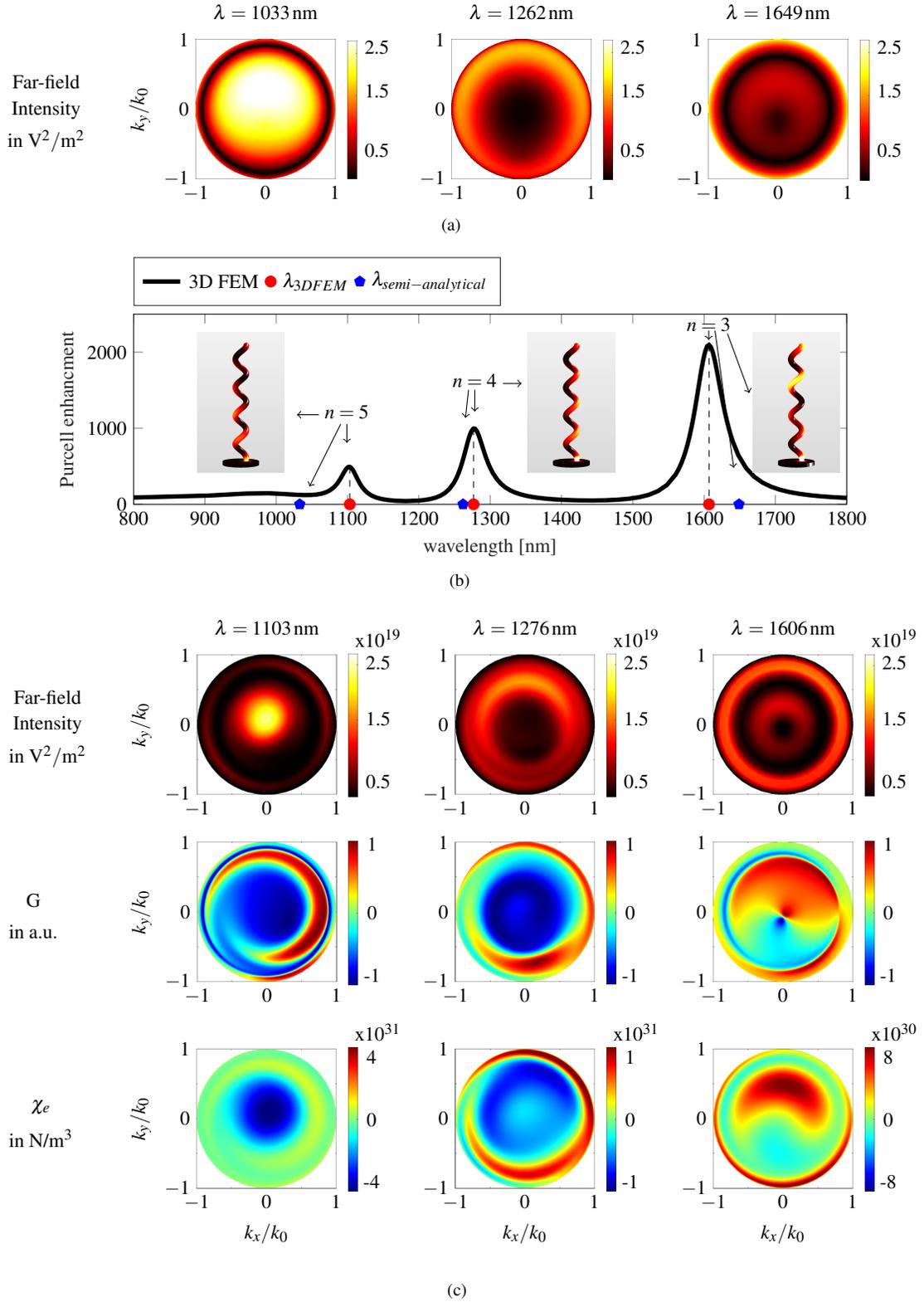

    \centering
      \hspace{-0.35cm}
      \subfigure[] 
      {\begin{tabular}[b]{lll}\input{Bilder/FIG_2_c/fig_2_c}\end{tabular}\label{fig:ErgebnisseSingleHelixa}}  
      \subfigure[] 
      {\input{Bilder/FIG_2_a/Purcelle_Single_Helix_w4_best}\label{fig:ErgebnisseSingleHelixb}}
      \subfigure[] 
      { \begin{tabular}[b]{lll}\input{Bilder/FIG_2_b/Fig_2_b}  \end{tabular}\label{fig:SingleHelixGundFernfeld}}  \hfill 
    \caption{Resonances and far-fields for the single helix with four pitches. 
    (a) Far-field computed with the semi-analytical model. 
    (b-c) Full 3D simulation results: (b) Purcell enhancement spectrum of a $z$-polarized dipole source at $x=60\,\text{nm}$, $8\,\text{nm}$ below the helix end cap. 
    Red (blue) dots indicate the resonance wavelengths from the full 3D (semi-analytical) results. 
    Insets show visualisations of the induced field intensity on the helix surface at the resonance wavelengths. 
    (c) Angular spectra of far field intensity, far-field degree of polarization $G$, and electric chirality density $\chi_e$.}
    \label{fig:ErgebnisseSingleHelix}
\end{figure*}
First, a semi-analytical extension of the design tool  of~\citet{hoflich2019resonant} (SA-DT) is used.
It approximates a single helix in vacuum by an one-dimensional linear rod resonator for calculation of the plasmonic Fabry-Perot modes dependent on the rod's radius, length and material~\cite{novotny_effective_2007}.
The overlap integral between a mode and a circularly polarized plane-wave excitation determines the total power that is transferred to the rod, when curved into the final helix geometry~\cite{Feichtner2017}.
The far-field radiation patterns of these modes can be assessed by varying the angle of incidence of the plane waves due to the  reciprocity of Maxwell's equations~\cite{Potton2004} (cf. Supporting Information for more detail).
Figure~\ref{fig:ErgebnisseSingleHelixa} shows far-field patterns acquired with the SA-DT at the resonance positions of the helix modes with orders $n = 3, 4$, and $5$ where $n$ equals the number of positions with vanishing surface charge, called nodes.
For $n=3$, nearly no emission occurs along the helix axis followed by a maximum for increasing emission angles, another minimum and a third maximum along the plane perpendicular to the helix.
The $n = 4$ mode exhibits one node per helix pitch corresponding to stacking the fundamental mode of the single pitch helix~\cite{Wozniak2018}.
In this case, all charge density maxima align along the $z$-axis which defines therefore also the net dipole orientation.
This results in the distinct dipole-like side emission observed in the far-field pattern.
Such an emission may be of interest for array/grating applications where the interaction of the elements leads to emerging lattice resonances~\cite{Petschulat2010, kravets2018}.
However, here we strive for the contrary, high directivity in $z$-direction to increase photon yield for any receiver/coupler for on-chip integration.
This is delivered by the $n=5$ mode.

For gaining a more complete understanding, the system is studied using full wave simulations using FEM and introducing a dipole emitter as the excitation source.
The power of a point emitter transferred to a resonator depends on the strength and overlap of the plasmonic resonator mode fields with the field of the emitter -- both classically and quantum mechanically~\cite{klimov2001}.
We investigate different positions and polarizations ($x$, $y$ and $z$ direction) of the dipole source.
The position is varied along the $x$-axis from $[0\,\text{nm}, 90\,\text{nm}]$ in $10\,\text{nm}$ steps, see sketch in  Fig.~\ref{fig:HelixAufbaua}. 
In the $z$-direction, the position of the dipole is located in the center of the $20\,\text{nm}$ hBN layer, $\Delta z = 8\,\text{nm}$ below the helix end cap. 
For each position and polarization, the Purcell enhancement is computed for the wavelength range $\lambda = [800\,\text{nm}, 1800\,\text{nm}]$ (see Supporting Information). 
The optimum coupling is achieved for a $z$-polarized dipole located exactly below the end cap of the helix at $x=60\,\text{nm}$ when the dipole is closest to the surface of the helix end cap.
The helix near-fields are strongest here and the dipole orientation matches the dominant near-field orientation normal to the metallic surface.

For the best coupling scenario, the Purcell enhancement $F$ in dependence on the wavelength is shown in Fig.~\ref{fig:ErgebnisseSingleHelixb} together with the mode field patterns as insets. 
The graph displays three clear resonances at $1606\,\text{nm}$, $1276\,\text{nm}$ and $1103\,\text{nm}$ with mode orders 3,4, and 5.
For comparison, the mode positions obtained by SA-DT are denoted by blue pentagons. 
Lower order modes feature higher Purcell enhancement due to their smaller mode volume, while the quality factors $Q$ are about 30 for all three resonances.
This is comparable to the case of straight rod antennas~\cite{encina_near-field_2009}.

Next, the far-field chirality properties $G$ and $\chi_e$ are studied. 
The first row in Fig.~\ref{fig:SingleHelixGundFernfeld} depicts far-field intensities obtained by the full-field model with dipole coupling for the three strong resonances. 
These closely resemble the results obtained from the semi-analytical approach but show an increased overall directivity, most probably caused by the influence of the substrate.
The angle-dependent portions of LCP and RCP light in the far-field are shown in the second row of Fig. ~\ref{fig:SingleHelixGundFernfeld}. 
For the $n = 3$ resonance at $1606\,\text{nm}$, a critical point of helicity is directly along the helix axis.
For the other two resonances at $1103\,\text{nm}$ and $1276\,\text{nm}$, RCP light dominates for emission along the helix axis and LCP in the outer part for oblique emission.
The electric chirality density $\chi_e$ is depicted in the third row of Fig. ~\ref{fig:SingleHelixGundFernfeld}.
This quantity provides a measure for the strength of chiroptical interaction which requires both field intensity and chirality.
The resonances at $1103\,\text{nm}$ and $1276\,\text{nm}$ exhibit a negative $\chi_e$ in the center corresponding to high-intensity RCP light emission for small angles.
However, the intensity distribution of the $n = 4$ resonance in the far-field is unfavorable for directional emission, as it decreases in the very center. 
In addition, $\chi_e$ shows large variations at higher angles making it hard to selectively address a specific handedness of light.
The same holds true for the $n = 3$ resonance at $1606\,\text{nm}$, where the far-field intensity corresponds to a higher order multipole emission and is therefore significantly weaker for emission along the antenna axis than for oblique emission.
Thus, only the $n=5$ resonance at $1103\,\text{nm}$ provides both, strongly directional emission of pure circularly polarized light.
The price for this is the lower Purcell enhancement compared to the lower order modes.
A possibility to achieve directional emission for these modes as well would therefore be highly desirable.

For this reason, we introduce in the following double helices that are C2 symmetric along the $z$-axis.

{\it Numerical Investigation of Double Helices} --
Next, helices with the same geometrical parameters but in a double helix configuration and differing number of pitches $p$ are studied. 
The excitation is realized by placing a dipole in the midpoint of the radius of the helix at the center of the hBN with its dipole moment oriented to point in the direction of both wires (here, $x$-axis), see Fig.~\ref{fig:HelixAufbaub}.
This in-plane alignment of the dipolar emitter is common for hBN color centers~\cite{Exarhos2017, Takashima2020}.
The central position was chosen as compromise between excitation efficiency and quenching~\cite{Marquier2017}. 

\begin{figure*}
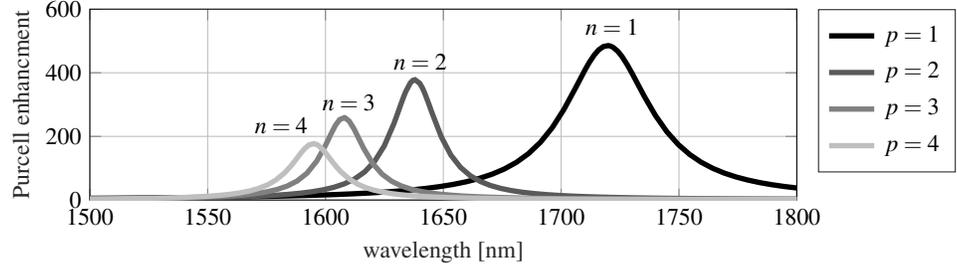
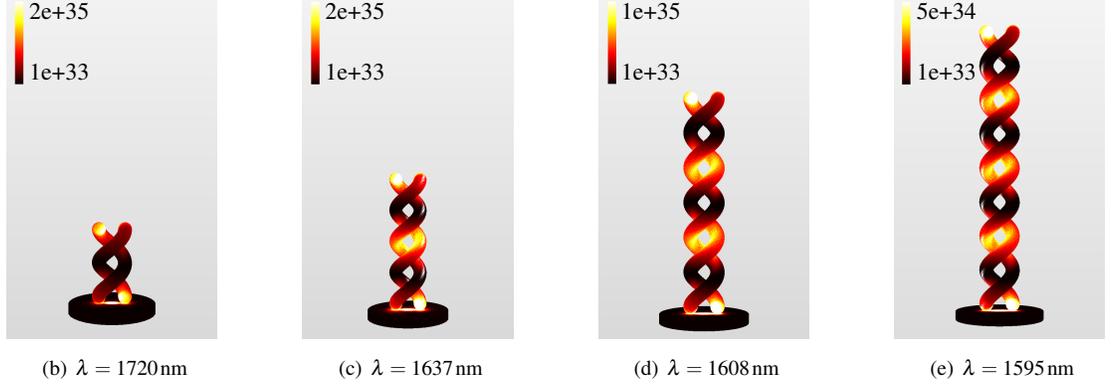
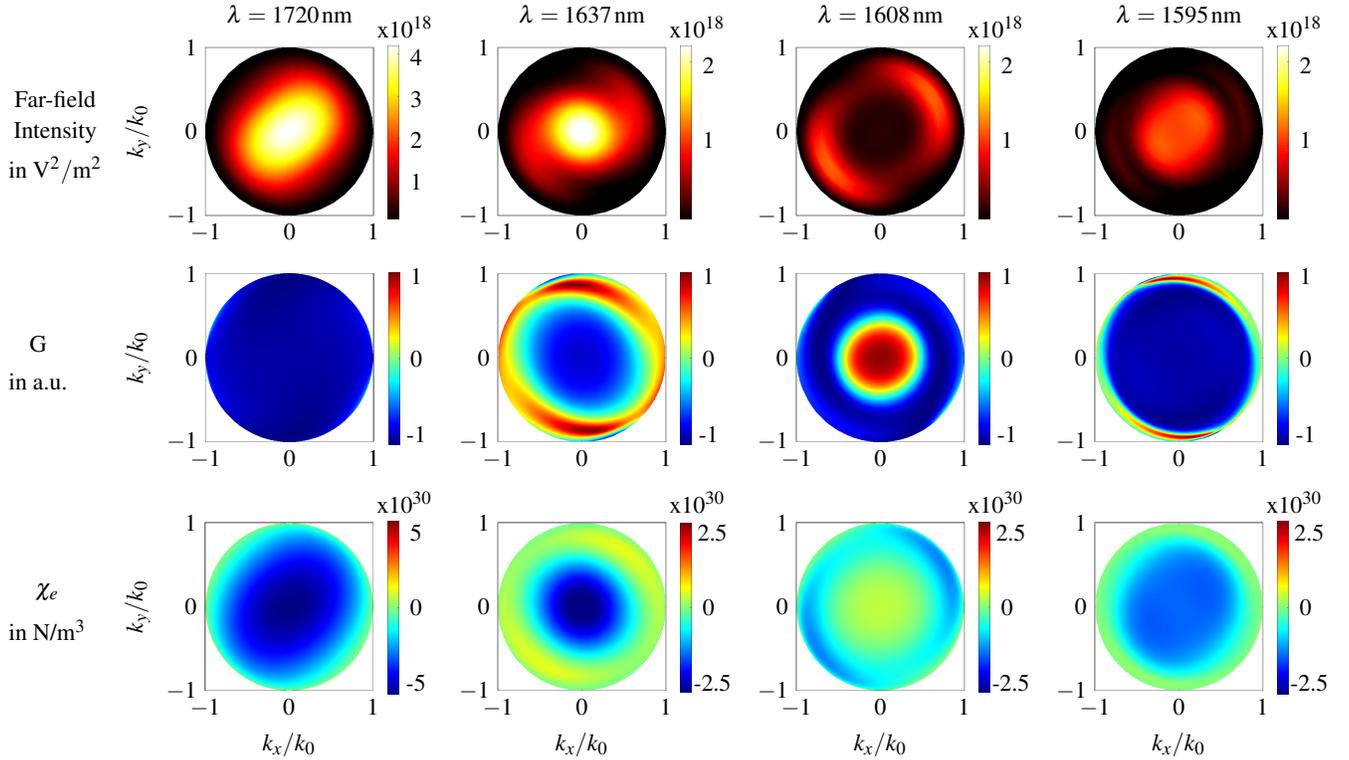

    \centering
    \subfigure[]{
%
%
\definecolor{mycolor1}{rgb}{0.00000,0.44700,0.74100}%
\definecolor{mycolor2}{rgb}{0.85000,0.32500,0.09800}%
\definecolor{mycolor3}{rgb}{0.92900,0.69400,0.12500}%
\definecolor{mycolor4}{rgb}{0.49400,0.18400,0.55600}%
\begin{tikzpicture}

\begin{axis}[%
width=3.7in,
height=1in,
at={(0.77in,0.474in)},
scale only axis,
xmin=1500,
xmax=1800,
xlabel style={font=\color{white!15!black}},
xlabel={wavelength [nm]},
ymin=0,
ymax=600,
ylabel style={font=\color{white!15!black}},
ylabel={Purcell enhancment},
axis background/.style={fill=white},
xmajorgrids,
ymajorgrids,
legend style={at={(1.03,1)}, anchor=north west, legend cell align=left, align=left, draw=white!15!black}
]
\addplot [color=black, line width=2.0pt]
  table[row sep=crcr]{%
1500	5.49249405476027\\
1511.15	5.99002195756115\\
1521.72	6.54176636747328\\
1531.76	7.15450612632306\\
1541.31	7.83576844195155\\
1550.43	8.59637695532767\\
1559.16	9.44821351284247\\
1567.52	10.4035623645657\\
1575.56	11.4811575617935\\
1583.29	12.6985423107392\\
1590.74	14.0801384379174\\
1597.94	15.6566739493771\\
1604.9	17.4616656244041\\
1611.65	19.5424616227103\\
1618.2	21.9525702963955\\
1624.57	24.7634596979259\\
1630.77	28.0618519969297\\
1636.82	31.9654870877817\\
1642.73	36.6217287431683\\
1648.51	42.2236859818683\\
1654.17	49.0273194011284\\
1659.72	57.3758042905947\\
1665.17	67.7340026541556\\
1670.53	80.737466927403\\
1675.8	97.2236146009805\\
1681.01	118.448095937153\\
1686.14	145.866214500698\\
1691.21	181.48550371322\\
1696.24	227.562511527644\\
1701.21	285.196581675503\\
1706.14	352.679719757184\\
1710	407.243367807521\\
1711.05	421.193587584593\\
1712	433.141839772866\\
1714	455.392786270141\\
1715.92	471.806465427099\\
1716	472.365069863832\\
1718	482.650490423626\\
1720	485.334772091034\\
1720.77	484.269766645013\\
1722	480.195010774751\\
1724	467.748393515257\\
1725.6	453.279776619368\\
1726	449.132831495641\\
1728	425.867714627096\\
1730	399.581356545466\\
1730.42	393.822902943508\\
1735.24	326.626451420738\\
1740.06	265.23179216318\\
1744.88	214.505225578033\\
1749.71	174.365188474283\\
1754.56	143.068569254106\\
1759.44	118.674922789526\\
1764.34	99.6009942721535\\
1769.28	84.4948024717176\\
1774.26	72.4134377893537\\
1779.29	62.6338665979649\\
1784.37	54.639827516558\\
1789.51	48.0318155480987\\
1794.72	42.5130780528318\\
1800	37.8686843615185\\
};
\addlegendentry{$p = 1$}

\addplot [color=black!65!white, line width=2.0pt]
  table[row sep=crcr]{%
1500	3.49461557530199\\
1511.23	3.98075192116302\\
1521.36	4.54733013719247\\
1530.57	5.20839117504657\\
1539.01	5.98257177446616\\
1546.79	6.89261392646478\\
1554.01	7.96914085451353\\
1560.75	9.25185518318974\\
1567.07	10.7914329386359\\
1573.04	12.6616619774461\\
1578.69	14.9521931836136\\
1584.07	17.7973371054194\\
1589.22	21.3870922485231\\
1594.16	25.9826996919463\\
1598.92	31.9807448328031\\
1603.53	39.9909818447276\\
1608.01	50.9521000933351\\
1612.37	66.32358049559\\
1616.63	88.4853917476027\\
1620.82	121.325907767939\\
1624.95	170.314873698117\\
1629.02	239.963369165968\\
1630	259.658424308375\\
1632	301.207417198822\\
1633.06	322.535027308948\\
1634	339.935100228435\\
1636	368.099177342812\\
1637.07	376.060906286522\\
1638	378.10569057142\\
1640	366.843236689502\\
1641.08	352.933660737577\\
1642	337.833162118843\\
1644	298.803770920526\\
1645.09	276.131098276565\\
1649.12	198.877726007817\\
1653.17	141.701248196141\\
1657.27	102.761437583403\\
1661.43	76.4989081978402\\
1665.65	58.500314531877\\
1669.97	45.7555417214788\\
1674.39	36.520486142759\\
1678.94	29.6388849518374\\
1683.62	24.4154086509631\\
1688.48	20.3465919977738\\
1693.53	17.1327967226594\\
1698.8	14.5541691729584\\
1704.32	12.4580339541537\\
1710.14	10.7304082010101\\
1716.29	9.29539500522814\\
1722.83	8.09096829715201\\
1729.82	7.07252655612261\\
1737.33	6.20604305859664\\
1745.45	5.46461195823688\\
1754.29	4.82731773163889\\
1763.97	4.27859108885138\\
1774.67	3.80466250335171\\
1786.59	3.3958841178603\\
1800	3.04429927710726\\
};
\addlegendentry{$p = 2$}

\addplot [color=black!50!white, line width=2.0pt ]
  table[row sep=crcr]{%
1500	2.9422906276838\\
1508.76	3.39145649632423\\
1516.81	3.92700670490891\\
1524.26	4.56843646626716\\
1531.2	5.34195883545743\\
1537.7	6.28232559537136\\
1543.81	7.43488281147702\\
1549.58	8.86310237246169\\
1555.07	10.6622024743744\\
1560.3	12.9568856778595\\
1565.32	15.9472508873359\\
1570.14	19.9117663251296\\
1574.79	25.2980980254283\\
1579.3	32.831747748823\\
1583.68	43.6693311415343\\
1587.95	59.7622833004072\\
1592.14	84.4056664978043\\
1596.25	122.20431231131\\
1600	171.966836565416\\
1600.3	176.49718675296\\
1602	202.749059769243\\
1604	231.862771577508\\
1604.3	235.700234856934\\
1606	252.76696415964\\
1608	259.002033883919\\
1608.27	258.526811540515\\
1610	248.401345656076\\
1612	224.777764737248\\
1612.22	221.676928522365\\
1614	194.995297029041\\
1616.16	162.663944534361\\
1620.1	114.150133666121\\
1624.06	80.9819248443157\\
1628.05	59.1244115119191\\
1632.08	44.5192806474241\\
1636.17	34.4560446362305\\
1640.32	27.3255396842403\\
1644.55	22.1141361571851\\
1648.89	18.1932169768871\\
1653.34	15.1878796872401\\
1657.94	12.8275970614556\\
1662.68	10.9541876610269\\
1667.61	9.4357816225473\\
1672.75	8.19099471715422\\
1678.13	7.15892786249379\\
1683.78	6.29535728643764\\
1689.75	5.56527865742157\\
1696.08	4.94426424361515\\
1702.83	4.41233689618455\\
1710.06	3.95479280542784\\
1717.86	3.55916219245498\\
1726.31	3.21693298960356\\
1735.54	2.92022373003187\\
1745.69	2.66356544238765\\
1756.95	2.44249556995424\\
1769.53	2.2543097654865\\
1783.75	2.09679715346477\\
1800	1.96958985988905\\
};
\addlegendentry{$p = 3$}

\addplot [color=black!25!white, line width=2.0pt]
  table[row sep=crcr]{%
1500	3.24345468810109\\
1508.29	3.76220013853214\\
1515.81	4.37953542494468\\
1522.69	5.11750496343922\\
1529.02	6.0034763674148\\
1534.89	7.07651641485098\\
1540.36	8.38629987678394\\
1545.5	10.0066568951411\\
1550.35	12.0341312862984\\
1554.95	14.6073327070893\\
1559.34	17.9309534932342\\
1563.53	22.2797891727034\\
1567.57	28.1143057038978\\
1571.47	36.079495533189\\
1575.26	47.1928255079441\\
1578.94	62.8183755692149\\
1582.55	84.8204646859096\\
1585	104.354445082634\\
1586.08	114.102151664737\\
1587	122.828471334535\\
1589	142.332996135911\\
1589.57	147.762203512283\\
1591	160.33282167333\\
1593	173.128863855431\\
1593.02	173.217412605423\\
1595	177.201943934946\\
1596.44	173.884400765919\\
1597	171.269567062923\\
1599	157.182354400891\\
1599.85	149.622037046061\\
1601	138.652345121001\\
1603	119.215427595871\\
1603.26	116.755950368259\\
1605	101.1332101081\\
1606.68	87.7467255354219\\
1610.12	65.8477487013316\\
1613.6	50.0872931934332\\
1617.13	38.8183837003311\\
1620.73	30.6483891990362\\
1624.41	24.6233275343366\\
1628.18	20.0965644875692\\
1632.07	16.6180142129705\\
1636.1	13.898409175946\\
1640.28	11.7444131744977\\
1644.65	10.007889396947\\
1649.23	8.59351644021958\\
1654.05	7.42940209359559\\
1659.16	6.45991997345367\\
1664.61	5.64573113873888\\
1670.44	4.95927403497716\\
1676.73	4.37603549955943\\
1683.55	3.87951681446878\\
1691	3.4558859826245\\
1699.22	3.09394457358371\\
1708.36	2.78633969878936\\
1718.61	2.52766349760214\\
1730.24	2.31399499481742\\
1743.61	2.14387543816175\\
1759.2	2.01907915598705\\
1777.67	1.94624422440588\\
1800	1.94116417914564\\
};
\addlegendentry{$p = 4$}
\end{axis}
\draw (8.9,3.5) node{$n=1$};
\draw (6.35,3.05) node{$n=2$};
\draw (5.4,2.5) node{$n=3$};
\draw (4.5,2.2) node{$n=4$};
\end{tikzpicture}%
\label{fig:DoubleHelicesa}}\\
    \hspace{2cm}
    \subfigure[$\lambda=1720\,\text{nm}$]{\input{Bilder/FIG_3_a_to_e/Feldbild_Double_Helices_w1}\label{fig:DoubleHelicesb}}\hspace{0.77cm}
    \subfigure[$\lambda=1637\,\text{nm}$]{\input{Bilder/FIG_3_a_to_e/Feldbild_Double_Helices_w2}\label{fig:DoubleHelicesc}}\hspace{0.77cm}
    \subfigure[$\lambda=1608\,\text{nm}$]{\input{Bilder/FIG_3_a_to_e/Feldbild_Double_Helices_w3}\label{fig:DoubleHelicesd}}\hspace{0.77cm}
    \subfigure[$\lambda=1595\,\text{nm}$]{\input{Bilder/FIG_3_a_to_e/Feldbild_Double_Helices_w4}\label{fig:DoubleHelicese}}\\
   \subfigure[]{\begin{tabular}[b]{llll}\input{Bilder/FIG_3_f/fig_3_f} \end{tabular}\label{fig:DoubleHelicesf}}
    \caption{(a) Purcell enhancement in dependency of the wavelength for a double helix with one, two, three, and four pitches $p$. 
    The intensity of the electric field strength for pitches $p$: (b) $p=1$, (c) $p=2$, (d) $p=3$ and (e) $p=4$. (f) Numerical results for the degree of polarization $G$, the far-field intensity, and electric chirality density $\chi_e$. }
    \label{fig:DoubleHelices}
\end{figure*}

Fig.~\ref{fig:DoubleHelicesa} displays the spectral dependency of the Purcell enhancement in the near-infrared range where one dominant resonance occurs at wavelengths of $\lambda=1720\,\text{nm}$ for one,  $\lambda=1637\,\text{nm}$ for two, $\lambda=1608\,\text{nm}$ for three, and $\lambda=1595\,\text{nm}$ for four pitches.
As can be seen in the surface intensity plots below, this is the resonance with one dipole per helix pitch, equaling a mode order of $n=1$ for one pitch up to mode order $n=4$ for four pitches.
The mode positions vary since the relative influence of the wire radius increases with decreasing wire length affecting the effective plasmon wavelength~\cite{novotny_effective_2007, hoflich2019resonant}. 

The strength of the Purcell enhancement decreases for increasing double helix pitches due to the increasing mode volume of the resonances.
Note, that the absolute Purcell enhancement values cannot be compared to the single helix results as distance and orientation between emitter and the helices differ.
The quality factor for the single pitch double helix is comparable to that of the single helix modes with $Q_{p=1} = 39$, while the double helices with 2, 3, and 4 pitches exhibit roughly doubled quality factors of $Q_{p=2} = 66$, $Q_{p=3} = 76$, and $Q_{p=4} = 66$.

Compared to the single helix geometry with 4 pitches where the $n=4$ resonance was at $\lambda=1276\,\text{nm}$ the near-field coupling between the helix arms causes a strong red-shift with the same mode order now appearing at $\lambda=1595\,\text{nm}$\cite{moradi_plasmon_2011}.
This antisymmetric coupling leads to effective dipole moments in the $x$-$y$-plane that radiate mainly in $z$-direction and interfere according to their relative phase difference. 
A simple model that explains the emission strength based of phase differences of dipoles stacked along the $z$-axis can be found in the Supporting Information.
Accordingly, the resulting net dipole moment has a high chance to be oriented in the $x$-$y$-plane leading to the desired strongly directional emission.
The first row of Fig.~\ref{fig:DoubleHelicesf} shows the far-field radiation patterns which indeed prove the directivity along the $z$-axis.
Only in the case of $p=3$ the highest intensity can be found in two side lobes and not within the center of the far-field due to destructive interference (cf. Supporting Information for more details).

The degree of circular polarization $G$ in the second row of Fig.~\ref{fig:DoubleHelicesf} shows that for all pitch numbers the RCP emission angles coincide with the strongest intensities in the far-field.
For one and four pitches, this RCP light emission occurs in a broad range of angles centered around the helix axis.
For two and three pitches, on the other hand, the RCP emission angles tend to be limited to emission along the antenna axis ($p = 2$) or rotationally symmetric oblique ($p = 3$).

The plots of the electric chirality density $\chi_e$ are displayed in the lower row of Fig.~\ref{fig:DoubleHelicesf}.
Both double helices with one and four pitches exhibit a large range of angles for the emission of RCP light, but with a significantly lower intensity for the latter case. 
While the double helix with three pitches lacks directional emission of RCP light, a very high directivity of RCP emission is obtained for the case of 2 pitches.
However, the largest $\chi_e$ values are found for the shortest helix, providing nearly perfect directivity and circularly polarized emission while being compact and easy to fabricate.

{\it Conclusion} --
Plasmonic resonators are an efficient tool to enhance light-matter interaction at the nano-scale and to alter the polarization state of the involved photons.
Plasmonic helices, in particular, are able to efficiently couple linearly polarized dipolar emitters to chiral radiation and vice versa. 
In this numerical study we demonstrated different far-field radiation properties in both directivity and polarization state of helical antennas driven by the same single emitter.
For assessing the underlying physics an extended semi-analytical design tool is employed, that evaluates helix geometries for resonance positions and qualitative far-field patterns.
Based on this a subsequent full-field modeling proved that a single emitter in a thin film such as hBN can be efficiently coupled to single and double helices.
Both antenna geometries provide strong Purcell enhancement in the telecom range, a very small footprint and act as a passive device that does not require electronic control such as phase shifters for polarization control~\cite{Moody2022}.
While the far-field emission angles of the examined single helices show a large variation depending on the mode order, the double helix provides highly directional emission already for a single pitch.
The versatility of emission patterns makes the helical antenna a potential building block for different applications such as chiral plasmonic phase gratings~\cite{lin_polarization_2013}, or sources of circularly polarized single-photons.
The latter relies on the high Purcell enhancements in combination with the fact that the emitted photons possess a well-defined spin and a high directivity.
In addition, the double helix can act as a device to couple chiral far-fields to the anti-symmetric mode of plasmonic two-wire waveguides~\cite{huang_impedance_2009} to realize a highly directional nanoscale polarization converter.
Finding optimal helical antenna configurations will speed up in the future by using the semi-analytical model for pre-screening large parameter spaces.
Subsequent full field simulations in combination with an optimization algorithm can quantitatively optimize the performance for specific applications. 
Therewith, we provide a route towards efficient implementation of plasmonic helices in various architectures for photonic quantum technologies.

\begin{acknowledgments}
Funded by the Deutsche Forschungsgemeinschaft (DFG, German Research Foundation) under Germany´s Excellence Strategy – The Berlin Mathematics Research Center MATH+ (EXC-2046/1, project ID: 390685689) and also by DFG under project ID HO5461/3-1 "chiralFEBID".
This project (20FUN05 SEQUME) has received funding from the EMPIR programme co-financed by the Participating States and from the European Union’s Horizon 2020 research and innovation programme.
This project has received funding from the German Federal Ministry of Education and Research (BMBF Forschungscampus  MODAL, project number 05M20ZBM).
\end{acknowledgments}

\section*{Data Availability Statement}
The input data files for the FEM simulations, for parameter scans,
and the simulation results of this study can be found in a corresponding open access data publication \cite{zenodo}.


\begin{thebibliography}{40}%
\makeatletter
\providecommand \@ifxundefined [1]{%
 \@ifx{#1\undefined}
}%
\providecommand \@ifnum [1]{%
 \ifnum #1\expandafter \@firstoftwo
 \else \expandafter \@secondoftwo
 \fi
}%
\providecommand \@ifx [1]{%
 \ifx #1\expandafter \@firstoftwo
 \else \expandafter \@secondoftwo
 \fi
}%
\providecommand \natexlab [1]{#1}%
\providecommand \enquote  [1]{``#1''}%
\providecommand \bibnamefont  [1]{#1}%
\providecommand \bibfnamefont [1]{#1}%
\providecommand \citenamefont [1]{#1}%
\providecommand \href@noop [0]{\@secondoftwo}%
\providecommand \href [0]{\begingroup \@sanitize@url \@href}%
\providecommand \@href[1]{\@@startlink{#1}\@@href}%
\providecommand \@@href[1]{\endgroup#1\@@endlink}%
\providecommand \@sanitize@url [0]{\catcode `\\12\catcode `\$12\catcode
  `\&12\catcode `\#12\catcode `\^12\catcode `\_12\catcode `\%12\relax}%
\providecommand \@@startlink[1]{}%
\providecommand \@@endlink[0]{}%
\providecommand \url  [0]{\begingroup\@sanitize@url \@url }%
\providecommand \@url [1]{\endgroup\@href {#1}{\urlprefix }}%
\providecommand \urlprefix  [0]{URL }%
\providecommand \Eprint [0]{\href }%
\providecommand \doibase [0]{http://dx.doi.org/}%
\providecommand \selectlanguage [0]{\@gobble}%
\providecommand \bibinfo  [0]{\@secondoftwo}%
\providecommand \bibfield  [0]{\@secondoftwo}%
\providecommand \translation [1]{[#1]}%
\providecommand \BibitemOpen [0]{}%
\providecommand \bibitemStop [0]{}%
\providecommand \bibitemNoStop [0]{.\EOS\space}%
\providecommand \EOS [0]{\spacefactor3000\relax}%
\providecommand \BibitemShut  [1]{\csname bibitem#1\endcsname}%
\let\auto@bib@innerbib\@empty
\bibitem [{\citenamefont {Flamini}\ and\ \citenamefont
  {Spagnolo}(2019)}]{Flamini2019}%
  \BibitemOpen
  \bibfield  {author} {\bibinfo {author} {\bibfnamefont {F.}~\bibnamefont
  {Flamini}}\ and\ \bibinfo {author} {\bibfnamefont {N.}~\bibnamefont
  {Spagnolo}},\ }\bibfield  {title} {\enquote {\bibinfo {title} {{Photonic
  quantum information processing : a review}},}\ }\href {\doibase
  10.1088/1361-6633/aad5b2} {\bibfield  {journal} {\bibinfo  {journal} {Rep.
  Prog. Phys.}\ }\textbf {\bibinfo {volume} {82}},\ \bibinfo {pages} {016001}
  (\bibinfo {year} {2019})}\BibitemShut {NoStop}%
\bibitem [{\citenamefont {Moody}\ \emph {et~al.}(2022)\citenamefont {Moody},
  \citenamefont {Sorger}, \citenamefont {Blumenthal}, \citenamefont
  {Juodawlkis}, \citenamefont {Loh}, \citenamefont {Sorace-Agaskar},
  \citenamefont {Jones}, \citenamefont {Balram}, \citenamefont {Matthews},
  \citenamefont {Laing}, \citenamefont {Davanco}, \citenamefont {Chang},
  \citenamefont {Bowers}, \citenamefont {Quack}, \citenamefont {Galland},
  \citenamefont {Aharonovich}, \citenamefont {Wolff}, \citenamefont {Schuck},
  \citenamefont {Sinclair}, \citenamefont {Lon{\v{c}}ar}, \citenamefont
  {Komljenovic}, \citenamefont {Weld}, \citenamefont {Mookherjea},
  \citenamefont {Buckley}, \citenamefont {Radulaski}, \citenamefont
  {Reitzenstein}, \citenamefont {Pingault}, \citenamefont {Machielse},
  \citenamefont {Mukhopadhyay}, \citenamefont {Akimov}, \citenamefont
  {Zheltikov}, \citenamefont {Agarwal}, \citenamefont {Srinivasan},
  \citenamefont {Lu}, \citenamefont {Tang}, \citenamefont {Jiang},
  \citenamefont {McKenna}, \citenamefont {Safavi-Naeini}, \citenamefont
  {Steinhauer}, \citenamefont {Elshaari}, \citenamefont {Zwiller},
  \citenamefont {Davids}, \citenamefont {Martinez}, \citenamefont {Gehl},
  \citenamefont {Chiaverini}, \citenamefont {Mehta}, \citenamefont {Romero},
  \citenamefont {Lingaraju}, \citenamefont {Weiner}, \citenamefont {Peace},
  \citenamefont {Cernansky}, \citenamefont {Lobino}, \citenamefont {Diamanti},
  \citenamefont {Vidarte},\ and\ \citenamefont {Camacho}}]{Moody2022}%
  \BibitemOpen
  \bibfield  {author} {\bibinfo {author} {\bibfnamefont {G.}~\bibnamefont
  {Moody}}, \bibinfo {author} {\bibfnamefont {V.~J.}\ \bibnamefont {Sorger}},
  \bibinfo {author} {\bibfnamefont {D.~J.}\ \bibnamefont {Blumenthal}},
  \bibinfo {author} {\bibfnamefont {P.~W.}\ \bibnamefont {Juodawlkis}},
  \bibinfo {author} {\bibfnamefont {W.}~\bibnamefont {Loh}}, \bibinfo {author}
  {\bibfnamefont {C.}~\bibnamefont {Sorace-Agaskar}}, \bibinfo {author}
  {\bibfnamefont {A.~E.}\ \bibnamefont {Jones}}, \bibinfo {author}
  {\bibfnamefont {K.~C.}\ \bibnamefont {Balram}}, \bibinfo {author}
  {\bibfnamefont {J.~C.~F.}\ \bibnamefont {Matthews}}, \bibinfo {author}
  {\bibfnamefont {A.}~\bibnamefont {Laing}}, \bibinfo {author} {\bibfnamefont
  {M.}~\bibnamefont {Davanco}}, \bibinfo {author} {\bibfnamefont
  {L.}~\bibnamefont {Chang}}, \bibinfo {author} {\bibfnamefont {J.~E.}\
  \bibnamefont {Bowers}}, \bibinfo {author} {\bibfnamefont {N.}~\bibnamefont
  {Quack}}, \bibinfo {author} {\bibfnamefont {C.}~\bibnamefont {Galland}},
  \bibinfo {author} {\bibfnamefont {I.}~\bibnamefont {Aharonovich}}, \bibinfo
  {author} {\bibfnamefont {M.~A.}\ \bibnamefont {Wolff}}, \bibinfo {author}
  {\bibfnamefont {C.}~\bibnamefont {Schuck}}, \bibinfo {author} {\bibfnamefont
  {N.}~\bibnamefont {Sinclair}}, \bibinfo {author} {\bibfnamefont
  {M.}~\bibnamefont {Lon{\v{c}}ar}}, \bibinfo {author} {\bibfnamefont
  {T.}~\bibnamefont {Komljenovic}}, \bibinfo {author} {\bibfnamefont
  {D.}~\bibnamefont {Weld}}, \bibinfo {author} {\bibfnamefont {S.}~\bibnamefont
  {Mookherjea}}, \bibinfo {author} {\bibfnamefont {S.}~\bibnamefont {Buckley}},
  \bibinfo {author} {\bibfnamefont {M.}~\bibnamefont {Radulaski}}, \bibinfo
  {author} {\bibfnamefont {S.}~\bibnamefont {Reitzenstein}}, \bibinfo {author}
  {\bibfnamefont {B.}~\bibnamefont {Pingault}}, \bibinfo {author}
  {\bibfnamefont {B.}~\bibnamefont {Machielse}}, \bibinfo {author}
  {\bibfnamefont {D.}~\bibnamefont {Mukhopadhyay}}, \bibinfo {author}
  {\bibfnamefont {A.}~\bibnamefont {Akimov}}, \bibinfo {author} {\bibfnamefont
  {A.}~\bibnamefont {Zheltikov}}, \bibinfo {author} {\bibfnamefont {G.~S.}\
  \bibnamefont {Agarwal}}, \bibinfo {author} {\bibfnamefont {K.}~\bibnamefont
  {Srinivasan}}, \bibinfo {author} {\bibfnamefont {J.}~\bibnamefont {Lu}},
  \bibinfo {author} {\bibfnamefont {H.~X.}\ \bibnamefont {Tang}}, \bibinfo
  {author} {\bibfnamefont {W.}~\bibnamefont {Jiang}}, \bibinfo {author}
  {\bibfnamefont {T.~P.}\ \bibnamefont {McKenna}}, \bibinfo {author}
  {\bibfnamefont {A.~H.}\ \bibnamefont {Safavi-Naeini}}, \bibinfo {author}
  {\bibfnamefont {S.}~\bibnamefont {Steinhauer}}, \bibinfo {author}
  {\bibfnamefont {A.~W.}\ \bibnamefont {Elshaari}}, \bibinfo {author}
  {\bibfnamefont {V.}~\bibnamefont {Zwiller}}, \bibinfo {author} {\bibfnamefont
  {P.~S.}\ \bibnamefont {Davids}}, \bibinfo {author} {\bibfnamefont
  {N.}~\bibnamefont {Martinez}}, \bibinfo {author} {\bibfnamefont
  {M.}~\bibnamefont {Gehl}}, \bibinfo {author} {\bibfnamefont {J.}~\bibnamefont
  {Chiaverini}}, \bibinfo {author} {\bibfnamefont {K.~K.}\ \bibnamefont
  {Mehta}}, \bibinfo {author} {\bibfnamefont {J.}~\bibnamefont {Romero}},
  \bibinfo {author} {\bibfnamefont {N.~B.}\ \bibnamefont {Lingaraju}}, \bibinfo
  {author} {\bibfnamefont {A.~M.}\ \bibnamefont {Weiner}}, \bibinfo {author}
  {\bibfnamefont {D.}~\bibnamefont {Peace}}, \bibinfo {author} {\bibfnamefont
  {R.}~\bibnamefont {Cernansky}}, \bibinfo {author} {\bibfnamefont
  {M.}~\bibnamefont {Lobino}}, \bibinfo {author} {\bibfnamefont
  {E.}~\bibnamefont {Diamanti}}, \bibinfo {author} {\bibfnamefont {L.~T.}\
  \bibnamefont {Vidarte}}, \ and\ \bibinfo {author} {\bibfnamefont {R.~M.}\
  \bibnamefont {Camacho}},\ }\bibfield  {title} {\enquote {\bibinfo {title}
  {{2022 Roadmap on integrated quantum photonics}},}\ }\href {\doibase
  10.1088/2515-7647/ac1ef4} {\bibfield  {journal} {\bibinfo  {journal} {J.
  Phys. Photonics}\ }\textbf {\bibinfo {volume} {4}},\ \bibinfo {pages} {12501}
  (\bibinfo {year} {2022})}\BibitemShut {NoStop}%
\bibitem [{\citenamefont {Maier}(2007)}]{Maier2007}%
  \BibitemOpen
  \bibfield  {author} {\bibinfo {author} {\bibfnamefont {S.~A.}\ \bibnamefont
  {Maier}},\ }\href@noop {} {\emph {\bibinfo {title} {{Plasmonics: Fundamentals
  and Applications}}}}\ (\bibinfo  {publisher} {Springer},\ \bibinfo {year}
  {2007})\BibitemShut {NoStop}%
\bibitem [{\citenamefont {Purcell}(1946)}]{purcell_spontaneous_1946}%
  \BibitemOpen
  \bibfield  {author} {\bibinfo {author} {\bibfnamefont {E.~M.}\ \bibnamefont
  {Purcell}},\ }\bibfield  {title} {\enquote {\bibinfo {title} {Spontaneous
  {Emission} {Probabilities} at {Radio} {Frequencies}},}\ }\href {\doibase
  10.1103/PhysRev.69.674.2} {\bibfield  {journal} {\bibinfo  {journal} {Phys.
  Rev.}\ }\textbf {\bibinfo {volume} {69}},\ \bibinfo {pages} {681} (\bibinfo
  {year} {1946})}\BibitemShut {NoStop}%
\bibitem [{\citenamefont {Kneipp}\ \emph {et~al.}(1997)\citenamefont {Kneipp},
  \citenamefont {Wang}, \citenamefont {Kneipp}, \citenamefont {Perelman},
  \citenamefont {Itzkan}, \citenamefont {Dasari},\ and\ \citenamefont
  {Feld}}]{Kneipp1997}%
  \BibitemOpen
  \bibfield  {author} {\bibinfo {author} {\bibfnamefont {K.}~\bibnamefont
  {Kneipp}}, \bibinfo {author} {\bibfnamefont {Y.}~\bibnamefont {Wang}},
  \bibinfo {author} {\bibfnamefont {H.}~\bibnamefont {Kneipp}}, \bibinfo
  {author} {\bibfnamefont {L.~T.}\ \bibnamefont {Perelman}}, \bibinfo {author}
  {\bibfnamefont {I.}~\bibnamefont {Itzkan}}, \bibinfo {author} {\bibfnamefont
  {R.~R.}\ \bibnamefont {Dasari}}, \ and\ \bibinfo {author} {\bibfnamefont
  {M.~S.}\ \bibnamefont {Feld}},\ }\bibfield  {title} {\enquote {\bibinfo
  {title} {Single molecule detection using surface-enhanced {R}aman scattering
  ({SERS})},}\ }\href@noop {} {\bibfield  {journal} {\bibinfo  {journal} {Phys.
  Rev. Lett.}\ }\textbf {\bibinfo {volume} {78}},\ \bibinfo {pages} {1667}
  (\bibinfo {year} {1997})}\BibitemShut {NoStop}%
\bibitem [{\citenamefont {K\"uhn}\ \emph {et~al.}(2006)\citenamefont {K\"uhn},
  \citenamefont {H\aa{}kanson}, \citenamefont {Rogobete},\ and\ \citenamefont
  {Sandoghdar}}]{Kuhn2006}%
  \BibitemOpen
  \bibfield  {author} {\bibinfo {author} {\bibfnamefont {S.}~\bibnamefont
  {K\"uhn}}, \bibinfo {author} {\bibfnamefont {U.}~\bibnamefont
  {H\aa{}kanson}}, \bibinfo {author} {\bibfnamefont {L.}~\bibnamefont
  {Rogobete}}, \ and\ \bibinfo {author} {\bibfnamefont {V.}~\bibnamefont
  {Sandoghdar}},\ }\bibfield  {title} {\enquote {\bibinfo {title} {Enhancement
  of single-molecule fluorescence using a gold nanoparticle as an optical
  nanoantenna},}\ }\href@noop {} {\bibfield  {journal} {\bibinfo  {journal}
  {Phys. Rev. Lett.}\ }\textbf {\bibinfo {volume} {97}},\ \bibinfo {pages}
  {017402} (\bibinfo {year} {2006})}\BibitemShut {NoStop}%
\bibitem [{\citenamefont {Kinkhabwala}\ \emph {et~al.}(2009)\citenamefont
  {Kinkhabwala}, \citenamefont {Yu}, \citenamefont {Fan}, \citenamefont
  {Avlasevich}, \citenamefont {Müllen},\ and\ \citenamefont
  {Moerner}}]{Kinkhabwala2009}%
  \BibitemOpen
  \bibfield  {author} {\bibinfo {author} {\bibfnamefont {A.}~\bibnamefont
  {Kinkhabwala}}, \bibinfo {author} {\bibfnamefont {Z.}~\bibnamefont {Yu}},
  \bibinfo {author} {\bibfnamefont {S.}~\bibnamefont {Fan}}, \bibinfo {author}
  {\bibfnamefont {Y.}~\bibnamefont {Avlasevich}}, \bibinfo {author}
  {\bibfnamefont {K.}~\bibnamefont {Müllen}}, \ and\ \bibinfo {author}
  {\bibfnamefont {W.~E.}\ \bibnamefont {Moerner}},\ }\bibfield  {title}
  {\enquote {\bibinfo {title} {{Large single-molecule fluorescence enhancements
  produced by a bowtie nanoantenna}},}\ }\href@noop {} {\bibfield  {journal}
  {\bibinfo  {journal} {Nat. Photonics}\ }\textbf {\bibinfo {volume} {3}},\
  \bibinfo {pages} {654} (\bibinfo {year} {2009})}\BibitemShut {NoStop}%
\bibitem [{\citenamefont {Novotny}\ and\ \citenamefont {van
  Hulst}(2011)}]{Novotny2011}%
  \BibitemOpen
  \bibfield  {author} {\bibinfo {author} {\bibfnamefont {L.}~\bibnamefont
  {Novotny}}\ and\ \bibinfo {author} {\bibfnamefont {N.}~\bibnamefont {van
  Hulst}},\ }\bibfield  {title} {\enquote {\bibinfo {title} {{Antennas for
  light}},}\ }\href@noop {} {\bibfield  {journal} {\bibinfo  {journal} {Nat.
  Photonics}\ }\textbf {\bibinfo {volume} {5}},\ \bibinfo {pages} {83}
  (\bibinfo {year} {2011})}\BibitemShut {NoStop}%
\bibitem [{\citenamefont {Biagioni}, \citenamefont {Huang},\ and\ \citenamefont
  {Hecht}(2012)}]{Biagioni2012}%
  \BibitemOpen
  \bibfield  {author} {\bibinfo {author} {\bibfnamefont {P.}~\bibnamefont
  {Biagioni}}, \bibinfo {author} {\bibfnamefont {J.-S.}\ \bibnamefont {Huang}},
  \ and\ \bibinfo {author} {\bibfnamefont {B.}~\bibnamefont {Hecht}},\
  }\bibfield  {title} {\enquote {\bibinfo {title} {{Nanoantennas for visible
  and infrared radiation}},}\ }\href@noop {} {\bibfield  {journal} {\bibinfo
  {journal} {Rep. Prog. Phys.}\ }\textbf {\bibinfo {volume} {75}},\ \bibinfo
  {pages} {024402} (\bibinfo {year} {2012})}\BibitemShut {NoStop}%
\bibitem [{\citenamefont {Koenderink}(2017)}]{Koenderink2017}%
  \BibitemOpen
  \bibfield  {author} {\bibinfo {author} {\bibfnamefont {A.~F.}\ \bibnamefont
  {Koenderink}},\ }\bibfield  {title} {\enquote {\bibinfo {title}
  {Single-photon nanoantennas},}\ }\href@noop {} {\bibfield  {journal}
  {\bibinfo  {journal} {ACS Photonics}\ }\textbf {\bibinfo {volume} {4}},\
  \bibinfo {pages} {710} (\bibinfo {year} {2017})}\BibitemShut {NoStop}%
\bibitem [{\citenamefont {Feichtner}, \citenamefont {Christiansen},\ and\
  \citenamefont {Hecht}(2017)}]{Feichtner2017}%
  \BibitemOpen
  \bibfield  {author} {\bibinfo {author} {\bibfnamefont {T.}~\bibnamefont
  {Feichtner}}, \bibinfo {author} {\bibfnamefont {S.}~\bibnamefont
  {Christiansen}}, \ and\ \bibinfo {author} {\bibfnamefont {B.}~\bibnamefont
  {Hecht}},\ }\bibfield  {title} {\enquote {\bibinfo {title} {{Mode Matching
  for Optical Antennas}},}\ }\href {\doibase 10.1103/PhysRevLett.119.217401} {\
  \textbf {\bibinfo {volume} {119}},\ \bibinfo {pages} {217401} (\bibinfo
  {year} {2017})}\BibitemShut {NoStop}%
\bibitem [{\citenamefont {Taminiau}\ \emph {et~al.}(2008)\citenamefont
  {Taminiau}, \citenamefont {Stefani}, \citenamefont {Segerink},\ and\
  \citenamefont {Hulst}}]{Taminiau2008}%
  \BibitemOpen
  \bibfield  {author} {\bibinfo {author} {\bibfnamefont {T.~H.}\ \bibnamefont
  {Taminiau}}, \bibinfo {author} {\bibfnamefont {F.~D.}\ \bibnamefont
  {Stefani}}, \bibinfo {author} {\bibfnamefont {F.~B.}\ \bibnamefont
  {Segerink}}, \ and\ \bibinfo {author} {\bibfnamefont {N.~F.~v.}\ \bibnamefont
  {Hulst}},\ }\bibfield  {title} {\enquote {\bibinfo {title} {{Optical antennas
  direct single-molecule emission}},}\ }\href@noop {} {\bibfield  {journal}
  {\bibinfo  {journal} {Nat. Photonics}\ }\textbf {\bibinfo {volume} {2}},\
  \bibinfo {pages} {234} (\bibinfo {year} {2008})}\BibitemShut {NoStop}%
\bibitem [{\citenamefont {Curto}\ \emph {et~al.}(2010)\citenamefont {Curto},
  \citenamefont {Volpe}, \citenamefont {Taminiau}, \citenamefont {Kreuzer},
  \citenamefont {Quidant},\ and\ \citenamefont {van Hulst}}]{Curto2010}%
  \BibitemOpen
  \bibfield  {author} {\bibinfo {author} {\bibfnamefont {A.~G.}\ \bibnamefont
  {Curto}}, \bibinfo {author} {\bibfnamefont {G.}~\bibnamefont {Volpe}},
  \bibinfo {author} {\bibfnamefont {T.~H.}\ \bibnamefont {Taminiau}}, \bibinfo
  {author} {\bibfnamefont {M.~P.}\ \bibnamefont {Kreuzer}}, \bibinfo {author}
  {\bibfnamefont {R.}~\bibnamefont {Quidant}}, \ and\ \bibinfo {author}
  {\bibfnamefont {N.~F.}\ \bibnamefont {van Hulst}},\ }\bibfield  {title}
  {\enquote {\bibinfo {title} {{Unidirectional Emission of a Quantum Dot
  Coupled to a Nanoantenna}},}\ }\href@noop {} {\bibfield  {journal} {\bibinfo
  {journal} {Science}\ }\textbf {\bibinfo {volume} {329}},\ \bibinfo {pages}
  {930} (\bibinfo {year} {2010})}\BibitemShut {NoStop}%
\bibitem [{\citenamefont {Gansel}\ \emph {et~al.}(2009)\citenamefont {Gansel},
  \citenamefont {Thiel}, \citenamefont {Rill}, \citenamefont {Decker},
  \citenamefont {Bade}, \citenamefont {Saile}, \citenamefont {von Freymann},
  \citenamefont {Linden},\ and\ \citenamefont {Wegener}}]{Gansel2009}%
  \BibitemOpen
  \bibfield  {author} {\bibinfo {author} {\bibfnamefont {J.~K.}\ \bibnamefont
  {Gansel}}, \bibinfo {author} {\bibfnamefont {M.}~\bibnamefont {Thiel}},
  \bibinfo {author} {\bibfnamefont {M.~S.}\ \bibnamefont {Rill}}, \bibinfo
  {author} {\bibfnamefont {M.}~\bibnamefont {Decker}}, \bibinfo {author}
  {\bibfnamefont {K.}~\bibnamefont {Bade}}, \bibinfo {author} {\bibfnamefont
  {V.}~\bibnamefont {Saile}}, \bibinfo {author} {\bibfnamefont
  {G.}~\bibnamefont {von Freymann}}, \bibinfo {author} {\bibfnamefont
  {S.}~\bibnamefont {Linden}}, \ and\ \bibinfo {author} {\bibfnamefont
  {M.}~\bibnamefont {Wegener}},\ }\bibfield  {title} {\enquote {\bibinfo
  {title} {{Gold helix photonic metamaterial as broadband circular
  polarizer.}}}\ }\href@noop {} {\bibfield  {journal} {\bibinfo  {journal}
  {Science}\ }\textbf {\bibinfo {volume} {325}},\ \bibinfo {pages} {1513}
  (\bibinfo {year} {2009})}\BibitemShut {NoStop}%
\bibitem [{\citenamefont {H{\"o}flich}\ \emph {et~al.}(2019)\citenamefont
  {H{\"o}flich}, \citenamefont {Feichtner}, \citenamefont {Hansj{\"u}rgen},
  \citenamefont {Haverkamp}, \citenamefont {Kollmann}, \citenamefont {Lienau},\
  and\ \citenamefont {Silies}}]{hoflich2019resonant}%
  \BibitemOpen
  \bibfield  {author} {\bibinfo {author} {\bibfnamefont {K.}~\bibnamefont
  {H{\"o}flich}}, \bibinfo {author} {\bibfnamefont {T.}~\bibnamefont
  {Feichtner}}, \bibinfo {author} {\bibfnamefont {E.}~\bibnamefont
  {Hansj{\"u}rgen}}, \bibinfo {author} {\bibfnamefont {C.}~\bibnamefont
  {Haverkamp}}, \bibinfo {author} {\bibfnamefont {H.}~\bibnamefont {Kollmann}},
  \bibinfo {author} {\bibfnamefont {C.}~\bibnamefont {Lienau}}, \ and\ \bibinfo
  {author} {\bibfnamefont {M.}~\bibnamefont {Silies}},\ }\bibfield  {title}
  {\enquote {\bibinfo {title} {Resonant behavior of a single plasmonic
  helix},}\ }\href@noop {} {\bibfield  {journal} {\bibinfo  {journal} {Optica}\
  }\textbf {\bibinfo {volume} {6}},\ \bibinfo {pages} {1098} (\bibinfo {year}
  {2019})}\BibitemShut {NoStop}%
\bibitem [{\citenamefont {Fernandez-Corbaton}, \citenamefont {Fruhnert},\ and\
  \citenamefont {Rockstuhl}(2016)}]{Fernandez-Corbaton2016}%
  \BibitemOpen
  \bibfield  {author} {\bibinfo {author} {\bibfnamefont {I.}~\bibnamefont
  {Fernandez-Corbaton}}, \bibinfo {author} {\bibfnamefont {M.}~\bibnamefont
  {Fruhnert}}, \ and\ \bibinfo {author} {\bibfnamefont {C.}~\bibnamefont
  {Rockstuhl}},\ }\bibfield  {title} {\enquote {\bibinfo {title} {Objects of
  maximum electromagnetic chirality},}\ }\href@noop {} {\bibfield  {journal}
  {\bibinfo  {journal} {Phys. Rev. X}\ }\textbf {\bibinfo {volume} {6}},\
  \bibinfo {pages} {031013} (\bibinfo {year} {2016})}\BibitemShut {NoStop}%
\bibitem [{\citenamefont {Santiago}\ \emph {et~al.}(2022)\citenamefont
  {Santiago}, \citenamefont {Hammerschmidt}, \citenamefont {Sachs},
  \citenamefont {Burger}, \citenamefont {Kwon}, \citenamefont {Kn{\"o}ller},
  \citenamefont {Arens}, \citenamefont {Fischer}, \citenamefont
  {Fernandez-Corbaton},\ and\ \citenamefont {Rockstuhl}}]{Garcia-Santiago2022}%
  \BibitemOpen
  \bibfield  {author} {\bibinfo {author} {\bibfnamefont {X.~G.}\ \bibnamefont
  {Santiago}}, \bibinfo {author} {\bibfnamefont {M.}~\bibnamefont
  {Hammerschmidt}}, \bibinfo {author} {\bibfnamefont {J.}~\bibnamefont
  {Sachs}}, \bibinfo {author} {\bibfnamefont {S.}~\bibnamefont {Burger}},
  \bibinfo {author} {\bibfnamefont {H.}~\bibnamefont {Kwon}}, \bibinfo {author}
  {\bibfnamefont {M.}~\bibnamefont {Kn{\"o}ller}}, \bibinfo {author}
  {\bibfnamefont {T.}~\bibnamefont {Arens}}, \bibinfo {author} {\bibfnamefont
  {P.}~\bibnamefont {Fischer}}, \bibinfo {author} {\bibfnamefont
  {I.}~\bibnamefont {Fernandez-Corbaton}}, \ and\ \bibinfo {author}
  {\bibfnamefont {C.}~\bibnamefont {Rockstuhl}},\ }\bibfield  {title} {\enquote
  {\bibinfo {title} {Toward maximally electromagnetically chiral scatterers at
  optical frequencies},}\ }\href {\doibase 10.1021/acsphotonics.1c01887}
  {\bibfield  {journal} {\bibinfo  {journal} {ACS Photonics}\ }\textbf
  {\bibinfo {volume} {9}},\ \bibinfo {pages} {1954} (\bibinfo {year}
  {2022})}\BibitemShut {NoStop}%
\bibitem [{\citenamefont {Wang}\ \emph {et~al.}(2019)\citenamefont {Wang},
  \citenamefont {Salut}, \citenamefont {Lu}, \citenamefont {Suarez},
  \citenamefont {Martin},\ and\ \citenamefont {Grosjean}}]{Wang2019}%
  \BibitemOpen
  \bibfield  {author} {\bibinfo {author} {\bibfnamefont {M.}~\bibnamefont
  {Wang}}, \bibinfo {author} {\bibfnamefont {R.}~\bibnamefont {Salut}},
  \bibinfo {author} {\bibfnamefont {H.}~\bibnamefont {Lu}}, \bibinfo {author}
  {\bibfnamefont {M.-A.}\ \bibnamefont {Suarez}}, \bibinfo {author}
  {\bibfnamefont {N.}~\bibnamefont {Martin}}, \ and\ \bibinfo {author}
  {\bibfnamefont {T.}~\bibnamefont {Grosjean}},\ }\bibfield  {title} {\enquote
  {\bibinfo {title} {{Subwavelength polarization optics via individual and
  coupled helical traveling-wave nanoantennas}},}\ }\href@noop {} {\bibfield
  {journal} {\bibinfo  {journal} {Light Sci. Appl.}\ }\textbf {\bibinfo
  {volume} {8}},\ \bibinfo {pages} {76} (\bibinfo {year} {2019})}\BibitemShut
  {NoStop}%
\bibitem [{\citenamefont {Aharonovich}, \citenamefont {Tetienne},\ and\
  \citenamefont {Toth}(2022)}]{Aharonovich2022}%
  \BibitemOpen
  \bibfield  {author} {\bibinfo {author} {\bibfnamefont {I.}~\bibnamefont
  {Aharonovich}}, \bibinfo {author} {\bibfnamefont {J.-P.}\ \bibnamefont
  {Tetienne}}, \ and\ \bibinfo {author} {\bibfnamefont {M.}~\bibnamefont
  {Toth}},\ }\bibfield  {title} {\enquote {\bibinfo {title} {Quantum emitters
  in hexagonal boron nitride},}\ }\href@noop {} {\bibfield  {journal} {\bibinfo
   {journal} {Nano Lett.}\ }\textbf {\bibinfo {volume} {22}},\ \bibinfo {pages}
  {9227} (\bibinfo {year} {2022})}\BibitemShut {NoStop}%
\bibitem [{\citenamefont {Johnson}\ and\ \citenamefont
  {Christy}(1972)}]{johnson1972optical}%
  \BibitemOpen
  \bibfield  {author} {\bibinfo {author} {\bibfnamefont {P.~B.}\ \bibnamefont
  {Johnson}}\ and\ \bibinfo {author} {\bibfnamefont {R.-W.}\ \bibnamefont
  {Christy}},\ }\bibfield  {title} {\enquote {\bibinfo {title} {Optical
  constants of the noble metals},}\ }\href@noop {} {\bibfield  {journal}
  {\bibinfo  {journal} {Phys. Rev. B}\ }\textbf {\bibinfo {volume} {6}},\
  \bibinfo {pages} {4370} (\bibinfo {year} {1972})}\BibitemShut {NoStop}%
\bibitem [{\citenamefont {Lee}\ \emph {et~al.}(2019)\citenamefont {Lee},
  \citenamefont {Jeong}, \citenamefont {Jung},\ and\ \citenamefont
  {Yee}}]{lee2019refractive}%
  \BibitemOpen
  \bibfield  {author} {\bibinfo {author} {\bibfnamefont {S.-Y.}\ \bibnamefont
  {Lee}}, \bibinfo {author} {\bibfnamefont {T.-Y.}\ \bibnamefont {Jeong}},
  \bibinfo {author} {\bibfnamefont {S.}~\bibnamefont {Jung}}, \ and\ \bibinfo
  {author} {\bibfnamefont {K.-J.}\ \bibnamefont {Yee}},\ }\bibfield  {title}
  {\enquote {\bibinfo {title} {Refractive index dispersion of hexagonal boron
  nitride in the visible and near-infrared},}\ }\href@noop {} {\bibfield
  {journal} {\bibinfo  {journal} {Phys. Status Solidi B}\ }\textbf {\bibinfo
  {volume} {256}},\ \bibinfo {pages} {1800417} (\bibinfo {year}
  {2019})}\BibitemShut {NoStop}%
\bibitem [{Sch(2017)}]{SchottBK7}%
  \BibitemOpen
  \href {https://refractiveindex.info/download/data/2017/schott_2017-01-20.pdf}
  {\enquote {\bibinfo {title} {{SCHOTT Zemax catalogue 2017-01-20b}},}\ }
  (\bibinfo {year} {2017})\BibitemShut {NoStop}%
\bibitem [{\citenamefont {Pomplun}\ \emph {et~al.}(2007)\citenamefont
  {Pomplun}, \citenamefont {Burger}, \citenamefont {Zschiedrich},\ and\
  \citenamefont {Schmidt}}]{pomplun2007adaptive}%
  \BibitemOpen
  \bibfield  {author} {\bibinfo {author} {\bibfnamefont {J.}~\bibnamefont
  {Pomplun}}, \bibinfo {author} {\bibfnamefont {S.}~\bibnamefont {Burger}},
  \bibinfo {author} {\bibfnamefont {L.}~\bibnamefont {Zschiedrich}}, \ and\
  \bibinfo {author} {\bibfnamefont {F.}~\bibnamefont {Schmidt}},\ }\bibfield
  {title} {\enquote {\bibinfo {title} {Adaptive finite element method for
  simulation of optical nano structures},}\ }\href@noop {} {\bibfield
  {journal} {\bibinfo  {journal} {Phys. Status Solidi B}\ }\textbf {\bibinfo
  {volume} {244}},\ \bibinfo {pages} {3419} (\bibinfo {year}
  {2007})}\BibitemShut {NoStop}%
\bibitem [{\citenamefont {Zschiedrich}\ \emph {et~al.}(2013)\citenamefont
  {Zschiedrich}, \citenamefont {Greiner}, \citenamefont {Burger},\ and\
  \citenamefont {Schmidt}}]{zschiedrich2013numerical}%
  \BibitemOpen
  \bibfield  {author} {\bibinfo {author} {\bibfnamefont {L.}~\bibnamefont
  {Zschiedrich}}, \bibinfo {author} {\bibfnamefont {H.~J.}\ \bibnamefont
  {Greiner}}, \bibinfo {author} {\bibfnamefont {S.}~\bibnamefont {Burger}}, \
  and\ \bibinfo {author} {\bibfnamefont {F.}~\bibnamefont {Schmidt}},\
  }\bibfield  {title} {\enquote {\bibinfo {title} {Numerical analysis of
  nanostructures for enhanced light extraction from {OLEDs}},}\ }\href
  {\doibase 10.1117/12.2001132} {\bibfield  {journal} {\bibinfo  {journal}
  {Proc. SPIE}\ }\textbf {\bibinfo {volume} {8641}},\ \bibinfo {pages} {86410B}
  (\bibinfo {year} {2013})}\BibitemShut {NoStop}%
\bibitem [{\citenamefont {Anger}, \citenamefont {Bharadwaj},\ and\
  \citenamefont {Novotny}(2006)}]{Anger2006}%
  \BibitemOpen
  \bibfield  {author} {\bibinfo {author} {\bibfnamefont {P.}~\bibnamefont
  {Anger}}, \bibinfo {author} {\bibfnamefont {P.}~\bibnamefont {Bharadwaj}}, \
  and\ \bibinfo {author} {\bibfnamefont {L.}~\bibnamefont {Novotny}},\
  }\bibfield  {title} {\enquote {\bibinfo {title} {{Enhancement and Quenching
  of Single-Molecule Fluorescence}},}\ }\href@noop {} {\bibfield  {journal}
  {\bibinfo  {journal} {Phys. Rev. Lett.}\ }\textbf {\bibinfo {volume} {96}},\
  \bibinfo {pages} {113002} (\bibinfo {year} {2006})}\BibitemShut {NoStop}%
\bibitem [{\citenamefont {Marquier}, \citenamefont {Sauvan},\ and\
  \citenamefont {Greffet}(2017)}]{Marquier2017}%
  \BibitemOpen
  \bibfield  {author} {\bibinfo {author} {\bibfnamefont {F.}~\bibnamefont
  {Marquier}}, \bibinfo {author} {\bibfnamefont {C.}~\bibnamefont {Sauvan}}, \
  and\ \bibinfo {author} {\bibfnamefont {J.-J.}\ \bibnamefont {Greffet}},\
  }\bibfield  {title} {\enquote {\bibinfo {title} {Revisiting quantum optics
  with surface plasmons and plasmonic resonators},}\ }\href@noop {} {\bibfield
  {journal} {\bibinfo  {journal} {ACS Photonics}\ }\textbf {\bibinfo {volume}
  {4}},\ \bibinfo {pages} {2091} (\bibinfo {year} {2017})}\BibitemShut
  {NoStop}%
\bibitem [{\citenamefont {Gutsche}\ \emph {et~al.}(2016)\citenamefont
  {Gutsche}, \citenamefont {Poulikakos}, \citenamefont {Hammerschmidt},
  \citenamefont {Burger},\ and\ \citenamefont {Schmidt}}]{gutsche2016time}%
  \BibitemOpen
  \bibfield  {author} {\bibinfo {author} {\bibfnamefont {P.}~\bibnamefont
  {Gutsche}}, \bibinfo {author} {\bibfnamefont {L.~V.}\ \bibnamefont
  {Poulikakos}}, \bibinfo {author} {\bibfnamefont {M.}~\bibnamefont
  {Hammerschmidt}}, \bibinfo {author} {\bibfnamefont {S.}~\bibnamefont
  {Burger}}, \ and\ \bibinfo {author} {\bibfnamefont {F.}~\bibnamefont
  {Schmidt}},\ }\bibfield  {title} {\enquote {\bibinfo {title} {Time-harmonic
  optical chirality in inhomogeneous space},}\ }\href {\doibase
  10.1117/12.2209551} {\bibfield  {journal} {\bibinfo  {journal} {Proc. SPIE}\
  }\textbf {\bibinfo {volume} {9756}},\ \bibinfo {pages} {97560X} (\bibinfo
  {year} {2016})}\BibitemShut {NoStop}%
\bibitem [{\citenamefont {Novotny}(2007)}]{novotny_effective_2007}%
  \BibitemOpen
  \bibfield  {author} {\bibinfo {author} {\bibfnamefont {L.}~\bibnamefont
  {Novotny}},\ }\bibfield  {title} {\enquote {\bibinfo {title} {Effective
  {Wavelength} {Scaling} for {Optical} {Antennas}},}\ }\href {\doibase
  10.1103/PhysRevLett.98.266802} {\bibfield  {journal} {\bibinfo  {journal}
  {Phys. Rev. Lett.}\ }\textbf {\bibinfo {volume} {98}},\ \bibinfo {pages}
  {266802} (\bibinfo {year} {2007})}\BibitemShut {NoStop}%
\bibitem [{\citenamefont {Potton}(2004)}]{Potton2004}%
  \BibitemOpen
  \bibfield  {author} {\bibinfo {author} {\bibfnamefont {R.~J.}\ \bibnamefont
  {Potton}},\ }\bibfield  {title} {\enquote {\bibinfo {title} {Reciprocity in
  optics},}\ }\href {\doibase 10.1088/0034-4885/67/5/R03} {\bibfield  {journal}
  {\bibinfo  {journal} {Rep. Prog. Phys.}\ }\textbf {\bibinfo {volume} {67}},\
  \bibinfo {pages} {717} (\bibinfo {year} {2004})}\BibitemShut {NoStop}%
\bibitem [{\citenamefont {Wo{\'{z}}niak}\ \emph {et~al.}(2018)\citenamefont
  {Wo{\'{z}}niak}, \citenamefont {{De Leon}}, \citenamefont {H{\"{o}}flich},
  \citenamefont {Haverkamp}, \citenamefont {Christiansen}, \citenamefont
  {Leuchs},\ and\ \citenamefont {Banzer}}]{Wozniak2018}%
  \BibitemOpen
  \bibfield  {author} {\bibinfo {author} {\bibfnamefont {P.}~\bibnamefont
  {Wo{\'{z}}niak}}, \bibinfo {author} {\bibfnamefont {I.}~\bibnamefont {{De
  Leon}}}, \bibinfo {author} {\bibfnamefont {K.}~\bibnamefont {H{\"{o}}flich}},
  \bibinfo {author} {\bibfnamefont {C.}~\bibnamefont {Haverkamp}}, \bibinfo
  {author} {\bibfnamefont {S.}~\bibnamefont {Christiansen}}, \bibinfo {author}
  {\bibfnamefont {G.}~\bibnamefont {Leuchs}}, \ and\ \bibinfo {author}
  {\bibfnamefont {P.}~\bibnamefont {Banzer}},\ }\bibfield  {title} {\enquote
  {\bibinfo {title} {{Chiroptical response of a single plasmonic nanohelix}},}\
  }\href@noop {} {\bibfield  {journal} {\bibinfo  {journal} {Opt. Express}\
  }\textbf {\bibinfo {volume} {26}},\ \bibinfo {pages} {1513} (\bibinfo {year}
  {2018})}\BibitemShut {NoStop}%
\bibitem [{\citenamefont {Petschulat}\ \emph {et~al.}(2010)\citenamefont
  {Petschulat}, \citenamefont {Cialla}, \citenamefont {Janunts}, \citenamefont
  {Rockstuhl}, \citenamefont {Hübner}, \citenamefont {Möller}, \citenamefont
  {Schneidewind}, \citenamefont {Mattheis}, \citenamefont {Popp}, \citenamefont
  {Tünnermann}, \citenamefont {Lederer},\ and\ \citenamefont
  {Pertsch}}]{Petschulat2010}%
  \BibitemOpen
  \bibfield  {author} {\bibinfo {author} {\bibfnamefont {J.}~\bibnamefont
  {Petschulat}}, \bibinfo {author} {\bibfnamefont {D.}~\bibnamefont {Cialla}},
  \bibinfo {author} {\bibfnamefont {N.}~\bibnamefont {Janunts}}, \bibinfo
  {author} {\bibfnamefont {C.}~\bibnamefont {Rockstuhl}}, \bibinfo {author}
  {\bibfnamefont {U.}~\bibnamefont {Hübner}}, \bibinfo {author} {\bibfnamefont
  {R.}~\bibnamefont {Möller}}, \bibinfo {author} {\bibfnamefont
  {H.}~\bibnamefont {Schneidewind}}, \bibinfo {author} {\bibfnamefont
  {R.}~\bibnamefont {Mattheis}}, \bibinfo {author} {\bibfnamefont
  {J.}~\bibnamefont {Popp}}, \bibinfo {author} {\bibfnamefont {A.}~\bibnamefont
  {Tünnermann}}, \bibinfo {author} {\bibfnamefont {F.}~\bibnamefont
  {Lederer}}, \ and\ \bibinfo {author} {\bibfnamefont {T.}~\bibnamefont
  {Pertsch}},\ }\bibfield  {title} {\enquote {\bibinfo {title} {{Doubly
  resonant optical nanoantenna arrays for polarization resolved measurements of
  surface-enhanced Raman scattering}},}\ }\href@noop {} {\bibfield  {journal}
  {\bibinfo  {journal} {Opt. Express}\ }\textbf {\bibinfo {volume} {18}},\
  \bibinfo {pages} {4184} (\bibinfo {year} {2010})}\BibitemShut {NoStop}%
\bibitem [{\citenamefont {Kravets}\ \emph {et~al.}(2018)\citenamefont
  {Kravets}, \citenamefont {Kabashin}, \citenamefont {Barnes},\ and\
  \citenamefont {Grigorenko}}]{kravets2018}%
  \BibitemOpen
  \bibfield  {author} {\bibinfo {author} {\bibfnamefont {V.~G.}\ \bibnamefont
  {Kravets}}, \bibinfo {author} {\bibfnamefont {A.~V.}\ \bibnamefont
  {Kabashin}}, \bibinfo {author} {\bibfnamefont {W.~L.}\ \bibnamefont
  {Barnes}}, \ and\ \bibinfo {author} {\bibfnamefont {A.~N.}\ \bibnamefont
  {Grigorenko}},\ }\bibfield  {title} {\enquote {\bibinfo {title} {Plasmonic
  {Surface} {Lattice} {Resonances}: {A} {Review} of {Properties} and
  {Applications}},}\ }\href {\doibase 10.1021/acs.chemrev.8b00243} {\bibfield
  {journal} {\bibinfo  {journal} {Chem. Rev.}\ }\textbf {\bibinfo {volume}
  {118}},\ \bibinfo {pages} {5912} (\bibinfo {year} {2018})}\BibitemShut
  {NoStop}%
\bibitem [{\citenamefont {Klimov}, \citenamefont {Ducloy},\ and\ \citenamefont
  {Letokhov}(2001)}]{klimov2001}%
  \BibitemOpen
  \bibfield  {author} {\bibinfo {author} {\bibfnamefont {V.~V.}\ \bibnamefont
  {Klimov}}, \bibinfo {author} {\bibfnamefont {M.}~\bibnamefont {Ducloy}}, \
  and\ \bibinfo {author} {\bibfnamefont {V.~S.}\ \bibnamefont {Letokhov}},\
  }\bibfield  {title} {\enquote {\bibinfo {title} {Spontaneous emission of an
  atom in the presence of nanobodies},}\ }\href {\doibase
  10.1070/QE2001v031n07ABEH002007} {\bibfield  {journal} {\bibinfo  {journal}
  {Quantum Electron.}\ }\textbf {\bibinfo {volume} {31}},\ \bibinfo {pages}
  {569} (\bibinfo {year} {2001})}\BibitemShut {NoStop}%
\bibitem [{\citenamefont {Encina}, \citenamefont {Perassi},\ and\ \citenamefont
  {Coronado}(2009)}]{encina_near-field_2009}%
  \BibitemOpen
  \bibfield  {author} {\bibinfo {author} {\bibfnamefont {E.~R.}\ \bibnamefont
  {Encina}}, \bibinfo {author} {\bibfnamefont {E.~M.}\ \bibnamefont {Perassi}},
  \ and\ \bibinfo {author} {\bibfnamefont {E.~A.}\ \bibnamefont {Coronado}},\
  }\bibfield  {title} {\enquote {\bibinfo {title} {Near-{Field} {Enhancement}
  of {Multipole} {Plasmon} {Resonances} in {Ag} and {Au} {Nanowires}},}\ }\href
  {\doibase 10.1021/jp811089a} {\bibfield  {journal} {\bibinfo  {journal} {J.
  Phys. Chem. A}\ }\textbf {\bibinfo {volume} {113}},\ \bibinfo {pages} {4489}
  (\bibinfo {year} {2009})}\BibitemShut {NoStop}%
\bibitem [{\citenamefont {Exarhos}\ \emph {et~al.}(2017)\citenamefont
  {Exarhos}, \citenamefont {Hopper}, \citenamefont {Grote}, \citenamefont
  {Alkauskas},\ and\ \citenamefont {Bassett}}]{Exarhos2017}%
  \BibitemOpen
  \bibfield  {author} {\bibinfo {author} {\bibfnamefont {A.~L.}\ \bibnamefont
  {Exarhos}}, \bibinfo {author} {\bibfnamefont {D.~A.}\ \bibnamefont {Hopper}},
  \bibinfo {author} {\bibfnamefont {R.~R.}\ \bibnamefont {Grote}}, \bibinfo
  {author} {\bibfnamefont {A.}~\bibnamefont {Alkauskas}}, \ and\ \bibinfo
  {author} {\bibfnamefont {L.~C.}\ \bibnamefont {Bassett}},\ }\bibfield
  {title} {\enquote {\bibinfo {title} {{Optical Signatures of Quantum Emitters
  in Suspended Hexagonal Boron Nitride}},}\ }\href {\doibase
  10.1021/acsnano.7b00665} {\bibfield  {journal} {\bibinfo  {journal} {ACS
  Nano}\ }\textbf {\bibinfo {volume} {11}},\ \bibinfo {pages} {3328} (\bibinfo
  {year} {2017})}\BibitemShut {NoStop}%
\bibitem [{\citenamefont {Takashima}\ \emph {et~al.}(2020)\citenamefont
  {Takashima}, \citenamefont {Maruya}, \citenamefont {Ishihara}, \citenamefont
  {Tashima}, \citenamefont {Shimazaki}, \citenamefont {Schell}, \citenamefont
  {Tran}, \citenamefont {Aharonovich},\ and\ \citenamefont
  {Takeuchi}}]{Takashima2020}%
  \BibitemOpen
  \bibfield  {author} {\bibinfo {author} {\bibfnamefont {H.}~\bibnamefont
  {Takashima}}, \bibinfo {author} {\bibfnamefont {H.}~\bibnamefont {Maruya}},
  \bibinfo {author} {\bibfnamefont {K.}~\bibnamefont {Ishihara}}, \bibinfo
  {author} {\bibfnamefont {T.}~\bibnamefont {Tashima}}, \bibinfo {author}
  {\bibfnamefont {K.}~\bibnamefont {Shimazaki}}, \bibinfo {author}
  {\bibfnamefont {A.~W.}\ \bibnamefont {Schell}}, \bibinfo {author}
  {\bibfnamefont {T.~T.}\ \bibnamefont {Tran}}, \bibinfo {author}
  {\bibfnamefont {I.}~\bibnamefont {Aharonovich}}, \ and\ \bibinfo {author}
  {\bibfnamefont {S.}~\bibnamefont {Takeuchi}},\ }\bibfield  {title} {\enquote
  {\bibinfo {title} {{Determination of the Dipole Orientation of Single Defects
  in Hexagonal Boron Nitride}},}\ }\href {\doibase
  10.1021/acsphotonics.0c00405} {\bibfield  {journal} {\bibinfo  {journal} {ACS
  Photonics}\ }\textbf {\bibinfo {volume} {7}},\ \bibinfo {pages} {2056}
  (\bibinfo {year} {2020})}\BibitemShut {NoStop}%
\bibitem [{\citenamefont {Moradi}(2011)}]{moradi_plasmon_2011}%
  \BibitemOpen
  \bibfield  {author} {\bibinfo {author} {\bibfnamefont {A.}~\bibnamefont
  {Moradi}},\ }\bibfield  {title} {\enquote {\bibinfo {title} {Plasmon
  hybridization in parallel nano-wire systems},}\ }\href {\doibase
  10.1063/1.3595236} {\bibfield  {journal} {\bibinfo  {journal} {Phys.
  Plasmas}\ }\textbf {\bibinfo {volume} {18}},\ \bibinfo {pages} {064508}
  (\bibinfo {year} {2011})}\BibitemShut {NoStop}%
\bibitem [{\citenamefont {Lin}\ \emph {et~al.}(2013)\citenamefont {Lin},
  \citenamefont {Mueller}, \citenamefont {Wang}, \citenamefont {Yuan},
  \citenamefont {Antoniou}, \citenamefont {Yuan},\ and\ \citenamefont
  {Capasso}}]{lin_polarization_2013}%
  \BibitemOpen
  \bibfield  {author} {\bibinfo {author} {\bibfnamefont {J.}~\bibnamefont
  {Lin}}, \bibinfo {author} {\bibfnamefont {J.~P.~B.}\ \bibnamefont {Mueller}},
  \bibinfo {author} {\bibfnamefont {Q.}~\bibnamefont {Wang}}, \bibinfo {author}
  {\bibfnamefont {G.}~\bibnamefont {Yuan}}, \bibinfo {author} {\bibfnamefont
  {N.}~\bibnamefont {Antoniou}}, \bibinfo {author} {\bibfnamefont {X.-C.}\
  \bibnamefont {Yuan}}, \ and\ \bibinfo {author} {\bibfnamefont
  {F.}~\bibnamefont {Capasso}},\ }\bibfield  {title} {\enquote {\bibinfo
  {title} {Polarization-controlled tunable directional coupling of surface
  plasmon polaritons},}\ }\href {\doibase 10.1126/science.1233746} {\bibfield
  {journal} {\bibinfo  {journal} {Science}\ }\textbf {\bibinfo {volume}
  {340}},\ \bibinfo {pages} {331} (\bibinfo {year} {2013})}\BibitemShut
  {NoStop}%
\bibitem [{\citenamefont {Huang}\ \emph {et~al.}(2009)\citenamefont {Huang},
  \citenamefont {Feichtner}, \citenamefont {Biagioni},\ and\ \citenamefont
  {Hecht}}]{huang_impedance_2009}%
  \BibitemOpen
  \bibfield  {author} {\bibinfo {author} {\bibfnamefont {J.-S.}\ \bibnamefont
  {Huang}}, \bibinfo {author} {\bibfnamefont {T.}~\bibnamefont {Feichtner}},
  \bibinfo {author} {\bibfnamefont {P.}~\bibnamefont {Biagioni}}, \ and\
  \bibinfo {author} {\bibfnamefont {B.}~\bibnamefont {Hecht}},\ }\bibfield
  {title} {\enquote {\bibinfo {title} {Impedance {Matching} and {Emission}
  {Properties} of {Nanoantennas} in an {Optical} {Nanocircuit}},}\ }\href
  {\doibase 10.1021/nl803902t} {\bibfield  {journal} {\bibinfo  {journal} {Nano
  Lett.}\ }\textbf {\bibinfo {volume} {9}},\ \bibinfo {pages} {1897} (\bibinfo
  {year} {2009})}\BibitemShut {NoStop}%
\bibitem [{\citenamefont {Kuen}\ \emph {et~al.}(2024)\citenamefont {Kuen},
  \citenamefont {Löffler}, \citenamefont {Tsarapkin}, \citenamefont
  {Zschiedrich}, \citenamefont {Feichtner}, \citenamefont {Burger},\ and\
  \citenamefont {Höflich}}]{zenodo}%
  \BibitemOpen
  \bibfield  {author} {\bibinfo {author} {\bibfnamefont {L.}~\bibnamefont
  {Kuen}}, \bibinfo {author} {\bibfnamefont {L.}~\bibnamefont {Löffler}},
  \bibinfo {author} {\bibfnamefont {A.}~\bibnamefont {Tsarapkin}}, \bibinfo
  {author} {\bibfnamefont {L.}~\bibnamefont {Zschiedrich}}, \bibinfo {author}
  {\bibfnamefont {T.}~\bibnamefont {Feichtner}}, \bibinfo {author}
  {\bibfnamefont {S.}~\bibnamefont {Burger}}, \ and\ \bibinfo {author}
  {\bibfnamefont {K.}~\bibnamefont {Höflich}},\ }\bibfield  {title} {\enquote
  {\bibinfo {title} {Source code and simulation results: Chiral and directional
  optical emission from a dipole source coupled to a helical plasmonic
  antenna},}\ }\href@noop {} {\bibfield  {journal} {\bibinfo  {journal}
  {Zenodo}\ ,\ \bibinfo {pages} {DOI: 10.5281/zenodo.10598256}} (\bibinfo
  {year} {2024})}\BibitemShut {NoStop}%
\end{thebibliography}
%

\end{document}